\documentclass[11pt]{article}

\usepackage{amsmath,amssymb}
\usepackage{graphicx, citesort}

\newtheorem{theorem}{Theorem}[section]
\newtheorem{lemma}[theorem]{Lemma}
\newtheorem{corollary}[theorem]{Corollary}

\newtheorem{definition}[theorem]{Definition}

\def\tr{{\hbox{\rm Tr}}}

\def\R{{{\mathbb{R}}}} 
\def\E{{\hbox{\bf E}}}
\def\P{{\hbox{\bf P}}}

\def\eps{\varepsilon}
\def \endprf{\hfill {\vrule height6pt width6pt depth0pt}\medskip}
\def\emph#1{{\it #1}} \def\textbf#1{{\bf #1}}

\newcommand{\<}{\langle}
\renewcommand{\>}{\rangle}

\newcommand{\coding}{A}
\newcommand{\decoding}{Q}

\numberwithin{equation}{section}
\def \endprf{\hfill {\vrule height6pt width6pt depth0pt}\medskip}
\newenvironment{proof}{\noindent {\bf Proof} }{\endprf\par}

\numberwithin{equation}{section}

\begin{document}

\title{\vspace{-5mm} Highly Robust Error Correction by Convex Programming}

\author{Emmanuel J. Cand\`es and Paige A. Randall\\[2mm]
Applied and Computational Mathematics, Caltech, Pasadena, CA  91125}

\date{November 2006} 

\maketitle

\begin{abstract}
  This paper discusses a stylized communications problem where one
  wishes to transmit a real-valued signal $x \in \R^n$ (a block of $n$
  pieces of information) to a remote receiver.  We ask whether it is
  possible to transmit this information reliably when a fraction of
  the transmitted codeword is corrupted by arbitrary gross errors, and
  when in addition, all the entries of the codeword are contaminated
  by smaller errors (e.g. quantization errors).

  We show that if one encodes the information as $\coding x$ where
  $\coding \in \R^{m \times n}$ $(m \ge n)$ is a suitable coding
  matrix, there are two decoding schemes that allow the recovery of
  the block of $n$ pieces of information $x$ with nearly the same
  accuracy as if no gross errors occur upon transmission (or
  equivalently as if one has an oracle supplying perfect information
  about the sites and amplitudes of the gross errors). Moreover, both
  decoding strategies are very concrete and only involve solving
  simple convex optimization programs, either a linear program or a
  second-order cone program. We complement our study with numerical
  simulations showing that the encoder/decoder pair performs
  remarkably well.
\end{abstract}


{\bf Acknowledgments.} This work has been partially supported by
National Science Foundation grants ITR ACI-0204932 and CCFœôòó515362 and
by the 2006 Waterman Award (NSF). E.~C. would like to thank the Centre
Interfacultaire Bernoulli of the Ecole Polytechnique F\'ed\'erale de
Lausanne for hospitality during June and July 2006. These results were
presented at WavE 2006, Lausanne, Switzerland, July 2006. Thanks to Mike Wakin for his careful reading of the manuscript. 

\section{Introduction} 

This paper discusses a coding problem over the reals.  We wish to
transmit a block of $n$ real values---a vector $x \in \R^n$---to a
remote receiver.  A possible way to address this problem is to
communicate the codeword $\coding x$ where $\coding$ is an $m$ by $n$
coding matrix with $m \ge n$. Now a recurrent problem with real
communication or storage devices is that some portions of the
transmitted codeword may become corrupted; when this occurs, parts of
the received codeword are unreliable and may have nothing to do with
their original values. We represent this as receiving a distorted
codeword $y = \coding x + z_0$. The question is whether one can
recover the signal $x$ from the received data $y$.

It has recently been shown \cite{DecodingLP,CRTV} that one could
recover the information $x$ exactly---under suitable conditions on the
coding matrix $\coding$---provided that the fraction of corrupted
entries of $\coding x$ is not too large. In greater details,
\cite{DecodingLP} proved that if the corruption $z_0$ contains at most
a fixed fraction of nonzero entries, then the signal $x\in \R^n$ is
the unique solution of the minimum-$\ell_1$ approximation problem
\begin{equation}
  \label{eq:DecodingLP}
  \min_{\tilde x \in \R^n} \|y - \coding \tilde x\|_{\ell_1}.  
\end{equation}
What may appear as a surprise is the fact that this requires no
assumption whatsoever about the corruption pattern $z_0$ except that
it must be sparse. In particular, the decoding algorithm is provably
exact even though the entries of $z_0$---and thus of $y$ as well---may
be arbitrary large, for example.

While this is interesting, it may not be realistic to assume that
except for some gross errors, one is able to receive the values of $\coding
x$ with infinite precision. A better model would assume instead that
the receiver gets
\begin{equation}
  \label{eq:model}
  y = \coding x + z_0, \qquad z_0 = e + z, 
\end{equation}
where $e$ is a possibly sparse vector of gross errors and $z$ is a
vector of small errors affecting all the entries. In other words, one
is willing to assume that there are malicious errors affecting a
fraction of the entries of the transmitted codeword {\em and} in
addition, smaller errors affecting all the entries. For instance, one
could think of $z$ as some sort of quantization error which limits the
precision/resolution of the transmitted information.  In this more
practical scenario, we ask whether it is still possible to recover the
signal $x$ accurately?  The subject of this paper is to show that it
is in fact possible to recover the original signal with nearly the
same accuracy as if one had a perfect communication system in which no
gross errors occur upon transmission. Further, the recovery algorithms
are especially simple, very concrete and practical; they involve
solving very convenient convex optimization problems.

To understand the results of this paper in a more quantitative
fashion, suppose that we had a perfect channel in which no gross
errors ever occur; that is, we assume $e = 0$ in \eqref{eq:model}. Then
we would receive $y = \coding x + z$ and would reconstruct $x$ by the
method of least-squares which, assuming that $\coding$ has full rank,
takes the form
\begin{equation}
  \label{eq:LS}
  x^{\text{Ideal}} = (\coding^* \coding)^{-1} \coding^* y.    
\end{equation}
In this ideal situation, the reconstruction error would then obey
\begin{equation}
  \label{eq:LSerror}
  \|x^{\text{Ideal}} - x\|_{\ell_2} = 
\|(\coding^* \coding)^{-1} \coding^* z\|_{\ell_2}. 
\end{equation}
Suppose we design the coding matrix $\coding$ with orthonormal columns
so that $\coding^* \coding = I$.  Then we would obtain a
reconstruction error whose maximum size is just about that of $z$.  If
the smaller errors $z_i$ are i.i.d.~$N(0,\sigma^2)$, then the
mean-squared error (MSE) would obey
\[
\E \|x^{\text{Ideal}} - x\|_{\ell_2}^2 = \sigma^2 \tr((\coding^* \coding)^{-1}).
\] 
If $A^*A = I$, then the MSE is equal to $n \sigma^2$. 

The question then is, can one hope to do almost as well as this
optimal mean squared error without knowing $e$ or even the support of
$e$ in advance?  This paper shows that one can in fact do almost as
well by solving very simple convex programs. This holds for {\em all}
signals $x \in \R^n$ and {\em all} sparse gross errors no matter how
adversary. 

Two concrete decoding strategies are introduced: one based on
second-order cone programming (SOCP) in Section \ref{sec:socp}, and
another based on linear programming (LP) in Section \ref{sec:lp}. We
will discuss the differences between the SOCP and the LP decoders, and
then compare their empirical performances in Section
\ref{sec:simulations}.

\section{Decoding by Second-Order Cone Programming}
\label{sec:socp}

To recover the signal $x$ from the corrupted vector $y$
\eqref{eq:model} we propose solving the following optimization program:
\begin{align}
  \label{eq:P2}
  (P_2) \quad \min \|y - \coding \tilde x - \tilde z\|_{\ell_1}
  \quad \text{ subject to } \quad   \|\tilde z\|_{\ell_2} & \le \eps,\\
  \nonumber \coding^* \tilde z & = 0,
\end{align}
with variables $\tilde x \in \R^n$ and $\tilde z \in \R^m$. The
parameter $\eps$ above depends on the magnitude of the small
errors and shall be specified later. The program $(P_2)$ is equivalent
to
\begin{align}
\label{eq:P2a}
   \min  {\bf 1}^* \tilde u , \quad \text{ subject to } \quad  & 
-\tilde u \le y - \coding \tilde x - \tilde z \le \tilde u,\\
\nonumber & \|\tilde z\|_{\ell_2} \le \eps,\\
\nonumber & \coding^* \tilde z = 0,  
\end{align}
where we added the slack optimization variable $\tilde u \in \R^m$. 
In the above formulation, ${\bf 1}$ is a vector of ones and the vector
inequality $u \le v$ means componentwise, i.e., $u_i \le v_i$ for all
$i$.  The program \eqref{eq:P2a} is a second-order cone program
and as a result, $(P_2)$ can be solved efficiently using standard
optimization algorithms, see \cite{BoydBook}.

The first key point of this paper is that the SOCP decoder is highly
robust against imperfections in communication channels.  Here and
below, $V$ denotes the subspace spanned by the columns of $\coding$,
and $\decoding \in \R^{m \times (m-n)}$ is a matrix whose columns form
an orthobasis of $V^\perp$, the orthogonal complement to $V$.  Such a
matrix $\decoding$ is a kind of parity-check matrix since $\decoding^*
\coding = 0$. Applying $\decoding^*$ on both sides of \eqref{eq:model}
gives
\begin{equation}
  \label{eq:model2}
  \decoding^* y = \decoding^* e + \decoding^* z. 
\end{equation}
Now if we could somehow get an accurate estimate $\hat e$ of $e$ from
$\decoding^* y$, we could reconstruct $x$ by applying the method of
Least Squares to the vector $y$ corrected for the gross errors:
\begin{equation}
  \label{eq:correctLS}
  \hat x = (\coding^* \coding)^{-1} \coding^* (y - \hat e). 
\end{equation}
If $\hat e$ were very accurate, we would probably do very well.

The point is that under suitable conditions, $(P_2)$ provides such
accurate estimates. Introduce $\tilde e = y - \coding \tilde x -
\tilde z$, and observe the following equivalence:
\begin{equation}
  \label{eq:equiv}
\begin{array}{cclllcll}
  (P_2) & \Leftrightarrow & \min & \|\tilde e\|_{\ell_1}  & 
\Leftrightarrow & (P'_2) \quad & \min & \|\tilde e\|_{\ell_1}\\
  & & \text{subject to} &  \tilde e  = y - \coding \tilde x - \tilde z, & & &  \text{subject to} &  \|\decoding^*(y - \tilde e)\|_{\ell_2} \le \eps.\\
  & & &  \coding^* \tilde z = 0, \,\,\,
  \|\tilde z\|_{\ell_2} \le \eps, & & & & 
\end{array}
\end{equation}
We only need to argue about the second equivalence since the first is
immediate. Observe that the condition $\coding^* \tilde z = 0$
decomposes $y - \tilde e$ as the superposition of an arbitrary element
in $V$ (the vector $\coding \tilde x$) and of an element in $V^\perp$
(the vector $\tilde z$) whose Euclidean length is less than $\eps$. In
other words, $\tilde z = P_{V^\perp}(y - \tilde e)$ where $P_{V^\perp}
= QQ^*$ is the orthonormal projector onto $V^\perp$ so that the
problem is that of minimizing the $\ell_1$ norm of $\tilde e$ under
the constraint $\|P_{V^\perp} (y - \tilde e)\|_{\ell_2} \le \eps$. The
claim follows from the identity $\|P_{V^\perp} v\|_{\ell_2} =
\|\decoding^* v\|_{\ell_2}$ which holds for all $v \in \R^m$.

The equivalence between $(P_2)$ and $(P'_2)$ asserts that if $(\hat x,
\hat z)$ is solution to $(P_2)$, then $\hat e = y - \coding \hat x -
\hat z$ is solution to $(P'_2)$ and vice versa; if $\hat e$ is
solution to $(P'_2)$, then there is a unique way to write $y - \hat e$
as the sum $\coding \hat x + \hat z$ with $z \in V^\perp$, and the
pair $(\hat x, \hat z)$ is solution to $(P_2)$.  We note, and this is
important, that the solution $\hat x$ to $(P_2)$ is also given by the
corrected least squares formula \eqref{eq:correctLS}. Equally
important is to note that even though we use the matrix $Q$ to explain
the rationale behind the methodology, one should keep in mind that $Q$
does not play any special role in $(P_2)$.

The issue here is that if $\|P_{V^\perp} v\|_{\ell_2}$ is
approximately proportional to $\|v\|_{\ell_2}$ for all sparse vectors
$v \in \R^m$, then the solution $\hat e$ to $(P'_2)$ is close to $e$,
provided that $e$ is sufficiently sparse \cite{CRT2}.  Quantitatively
speaking, if $\eps$ is chosen so that $\|P_{V^\perp} z\|_{\ell_2} \le
\eps$, then $\|e - \hat e\|$ is less than a numerical constant times
$\eps$; that is, the reconstruction error is within the noise level.
The key concept underlying this theory is the so-called {\em
  restricted isometry property}.
\begin{definition}
  Define the isometry constant $\delta_k$ of a matrix $\Phi$ as the
  smallest number such that
\begin{equation}
  \label{eq:rip}
  (1-\delta_k) \|x\|^2_{\ell_2} \le \|\Phi x\|^2_{\ell_2} \le 
  (1+\delta_k) \|x\|^2_{\ell_2} 
\end{equation}
holds for all $k$-sparse vectors $x$ (a $k$-sparse vector has at most
$k$ nonzero entries).
\end{definition}

In the sequel, we shall be concerned with the isometry constants of
$A^*$ times a scalar. Since $AA^*$ is the orthogonal projection $P_V$
onto $V$, we will be thus interested in subspaces $V$ such that $P_V$
nearly acts as an isometry on sparse vectors.  Our first result states
that the SOCP decoder is provably accurate.

\begin{theorem}
\label{teo:P2decode}
Choose a coding matrix $\coding \in \R^{m\times n}$ with orthonormal
columns spanning $V$, and let $(\delta_k)$ be the isometry constants
of the rescaled matrix $\sqrt{\frac{m}{n}} \, A^*$.  Suppose $\|P_{V^\perp}
z\|_{\ell_2} \le \eps$. 
Then the solution $\hat x$ to $(P_2)$ obeys
\begin{equation}
  \label{eq:P2decode}  
  \|\hat x - x\|_{\ell_2} \le 
C_2 \cdot \frac{\eps}{\sqrt{1-\frac{n}{m}}}  + 
\|x^{\text{\em Ideal}}- x\|_{\ell_2}
\end{equation} 
for some numerical constant $C_2$ provided that the number $k$ of
gross errors obeys $\delta_{3k} + \frac{1}{2} \delta_{2k} <
\frac{1}{2}(\frac{m}{n}-1)$; $x^{\text{\em Ideal}}$ is the ideal
solution \eqref{eq:LS} one would get if no gross errors ever occurred
$(e = 0)$.

If the (orthonormal) columns of $A$ are selected uniformly at random,
then with probability at least $1 - O(e^{-\gamma(m-n)})$ for some
positive constant $\gamma$, the estimate \eqref{eq:P2decode} holds for
$k \asymp \rho \cdot m$, provided $\rho \le \rho^*(n/m)$, which is a
constant depending only $n/m$.
\end{theorem}

This theorem is of significant appeal because it says that the
reconstruction error is in some sense within a constant factor of the
ideal solution. Indeed, suppose all we know about $z$ is that
$\|z\|_{\ell_2} \le \eps$. Then $\|x^{\text{Ideal}} - x\|_{\ell_2} =
\|A^* z\|_{\ell_2}$ may be as large as $\eps$. Thus for $m = 2n$, say,
\eqref{eq:P2decode} asserts that {\em the reconstruction error is
  bounded by a constant times the ideal reconstruction error.}  In
addition, if one selects a coding matrix with random orthonormal
columns (one way of doing so is to sample $X \in \R^{m \times n}$ with
i.i.d.~$N(0,1)$ entries and orthonormalize the columns by means of the
QR factorization), then one can correct a positive fraction of
arbitrarily corrupted entries, in a near ideal fashion.

Note that in the case where there are no small errors ($z = 0$), the
decoding is exact since $\eps = 0$ and $x^{\text{Ideal}} = x$. Hence,
this generalizes earlier results \cite{DecodingLP}. We would like to
emphasize that there is nothing special about the fact that the
columns of $A$ are taken to be orthonormal in Theorem
\ref{teo:P2decode}. In fact, one could just as well obtain equivalent
statements for general matrices. Our assumption only allows us to
formulate simple and useful results.

While the previous result discussed arbitrary small errors, the next
is about stochastic errors.
\begin{corollary}
\label{cor:P2decode}
Suppose the small errors are i.i.d.~$N(0,\sigma^2)$ and set $\eps :=
\sqrt{(m-n) (1 + t)}\cdot \sigma$ for some fixed $t > 0$.  Then under
the same hypotheses about the restricted isometry constants of $A$ and
the number of gross errors as in Theorem \ref{teo:P2decode}, the
solution to $(P_2)$ obeys
 \begin{equation}
    \label{eq:P2decode2}
    \|\hat x - x\|^2_{\ell_2} \le  C'_2 \cdot m \cdot \sigma^2,
  \end{equation}
  for some numerical constant $C'_2$ with probability exceeding $1 -
  e^{-\gamma^2 (m-n)/2} - e^{-m/2}$ where $\gamma =
  \frac{\sqrt{1+2t}-1}{\sqrt{2}}$. In particular, this last statement
  holds with overwhelming probability if $A$ is chosen at random as in
  Theorem \ref{teo:P2decode}.
\end{corollary}
Suppose for instance that $m = 2n$ to make things concrete so that the
MSE of the ideal estimate is equal to $m/2 \cdot \sigma^2$. Then the
SOCP reconstruction is within a multiplicative factor $2C$ of the
ideal MSE. Our experiments show that in practice the constant is
small: e.g. when $m = 2n$, one can correct 15\% of arbitrary errors, and 
in the overwhelming majority of cases obtain a decoded vector
whose MSE is less than 3 times larger than the ideal MSE.

\section{Decoding by Linear Programming}
\label{sec:lp}

Another way to recover the signal $x$ from the corrupted vector $y$
\eqref{eq:model} is by linear programming:
\begin{align}
  \label{eq:Pinfty}
  (P_\infty) \quad  \min \|y - \coding \tilde x - \tilde z\|_{\ell_1} 
\quad \text{ subject to } \quad   \|\tilde z\|_{\ell_\infty} & \le \lambda,\\
  \nonumber A^* \tilde z & = 0,
\end{align}
with variables $\tilde x \in \R^n$ and $\tilde z \in \R^m$.  As is
well known, the program $(P_\infty)$ may also be re-expressed as a
linear program by introducing slack variables just as in $(P_2)$; we
omit the standard details. As with $(P_2)$, the parameter $\lambda$
here is related to the size of the small errors and will be discussed
shortly. In the sequel, we shall also be interested in the more
general formulation of $(P_\infty)$
\begin{align}
  \label{eq:Pinftys}
 \|y - A \tilde x - z\|_{\ell_1} \quad \text{ subject to } \quad
  |\tilde z|_i & \le \lambda_i, \,\,\, 1 \le i \le m,\\
 \nonumber  A^* \tilde z & = 0,
\end{align}
which gives additional flexibility for adjusting the thresholds
$\lambda_1, \lambda_2, \ldots, \lambda_m$ to the noise level.

The same arguments as before prove that $(P_\infty)$ is equivalent to
\begin{equation}
  \label{eq:nPinfty}
  (P'_\infty) \quad  \min \|\tilde e\|_{\ell_1} 
  \quad \text{ subject to } \quad   
\|Q Q^*(y - \tilde e) \|_{\ell_\infty}  \le \lambda,  
\end{equation}
where we recall that $P_{V^\perp} = QQ^*$ is the orthonormal projector
onto $V^\perp$ ($V$ is the column space of $A$); that is, if $\hat e$
is solution to $(P'_\infty)$, then there is a unique decomposition $y
- \hat e = A \hat x + \hat z$ where $A^* \hat z = 0$ and $(\hat x,
\hat z)$ is solution to $(P_\infty)$. The converse is also
true. Similarly, the more general program \eqref{eq:Pinftys} is
equivalent to minimizing the $\ell_1$ norm of $\tilde e$ under the
constraint $|P_{V^\perp}(y - \tilde e)|_i \le \lambda_i$, $1 \le i \le m$.

In statistics, the estimator $\hat e$ solution to $(P'_\infty)$ is known as
the {\em Dantzig selector} \cite{Dantzig}. It was originally
introduced to estimate the vector $e$ from the data $y'$ and the model 
\begin{equation}
  \label{eq:stat-model}
  y' = Q^* e + z'
\end{equation}
where $z'$ is a vector of stochastic errors, e.g.~independent
mean-zero Gaussian random variables. The connection with our problem
is clear since applying the parity-check matrix $Q^*$ on both sides of
\eqref{eq:model} gives 
\[
Q^* y = Q^* e + Q^* z
\]
as before. If $z$ is stochastic noise, we can use the Dantzig
selector to recover $e$ from $Q^* y$. Moreover, available statistical
theory asserts that if $Q^*$ obeys nice restricted isometry properties
and $e$ is sufficiently sparse just as before, then this estimation
procedure is extremely accurate and in some sense optimal.

It remains to discuss how one should specify the parameter $\lambda$
in \eqref{eq:Pinfty}-\eqref{eq:nPinfty} which is easy. Suppose the small
errors are stochastic. Then we fix $\lambda$ so that the true vector
$e$ is feasible for $(P'_\infty)$ with very high probability; i.e. we
adjust $\lambda$ so that
\[
\|P_{V^\perp}(y - e)\|_{\ell_\infty} = \|P_{V^\perp} z\|_{\ell_\infty} \le \lambda  
\]
with high probability. In the more general formulation, the thresholds
are adjusted so that $\sup_{1\le i \le m} |P_{V^\perp} z|_i/\lambda_i \le 1$
with high probability.

The main result of this section is that the LP decoder is also
provably accurate.
\begin{theorem}
\label{teo:Pinftydecode}
Choose a coding matrix $\coding \in \R^{m\times n}$ with orthonormal
columns spanning $V$, and let $(\delta_k)$ be the isometry constants
of the rescaled matrix $\sqrt{\frac{m}{n}} \, A^*$.  Suppose
$\|P_{V^\perp} z\|_{\ell_\infty} \le \lambda$. Then the solution $\hat
x$ to $(P_\infty)$ obeys
\begin{equation}
  \label{eq:Pinftydecode}  
  \|\hat x - x\|_{\ell_2} \le 
C_1 \sqrt{k} \cdot \frac{\lambda}{1-\frac{n}{m}}  + 
\|x^{\text{\em Ideal}} - x\|_{\ell_2}
\end{equation} 
for some numerical constant $C_1$ provided that the number $k$ of gross
errors obeys $\delta_{3k} + \delta_{2k} < \frac{m}{n}-1$;
$x^{\text{\em Ideal}}$ is the ideal solution \eqref{eq:LS} one would
get if no gross errors ever occurred.

If the (orthonormal) columns of $A$ are selected uniformly at random,
then with probability at least $1 - O(e^{-\gamma(m-n)})$ for some
positive constant $\gamma$, the estimate \eqref{eq:Pinftydecode} holds
for $k \asymp \rho \cdot m$, provided $\rho \le \rho^*(n/m)$.
\end{theorem}

In effect, the LP decoder efficiently corrects a positive fraction of
arbitrarily corrupted entries.  Again, when there are no small errors
($z = 0$), the decoding is exact. (Also and just as before, there is
nothing special about the fact that the columns of $A$ are taken to be
orthonormal.) We now consider the interesting case in which the small
errors are stochastic. Below, we conveniently adjust the thresholds
$\lambda_j$ so that the true vector $e$ is feasible with high
probability, see Section \ref{sec:prooflp} for details.

\begin{corollary}
\label{cor:Pinftydecode}
Choose a coding matrix $A$ with (orthonormal) columns selected
uniformly at random and suppose the small errors are
i.i.d.~$N(0,\sigma^2)$. Fix
$$
\lambda_i = \sqrt{2\log m} \cdot \sqrt{1-\|A_{i, \cdot}\|^2_{\ell_2}}
\cdot \sigma
$$
in \eqref{eq:Pinftys}, where $\|A_{i, \cdot}\|_{\ell_2} = (\sum_{1
    \le j \le n} A^2_{i,j})^{1/2}$ is the $\ell_2$
  norm of the $i$th row.  Then if the number $k$ of gross errors is no
  more than a fraction of $m$ as in Theorem \ref{teo:Pinftydecode},
  the solution $\hat x$ obeys
 \begin{equation}
    \label{eq:Pinftydecode2}
    \|\hat x - x\|^2_{\ell_2} \le  [1 + C'_1 s]^2 \cdot  
    \|x^{\text{\em Ideal}}- x\|^2_{\ell_2} \qquad  s^2 =   
    \frac{k}{m} \cdot \frac{\log m}{\frac{n}{m}(1-\frac{n}{m})},  
  \end{equation}
  with very large probability, where $C'_1$ is some numerical
  constant. In effect, $\|\hat x - x\|^2_{\ell_2}$ is bounded by just
  about $[1 + C'_1 s]^2 \cdot n\sigma^2$ since $\|x^{\text{\em Ideal}}
  - x\|^2_{\ell_2}$ is distributed as $\sigma^2$ times a chi-square
  with $n$ degrees of freedom, and is tightly concentrated around
  $n\sigma^2$.
\end{corollary}
Recall that the MSE  is equal to $n \sigma^2$ when there are no
gross errors and, therefore, this last result asserts that {\em the
  reconstruction error is bounded by a constant times the ideal
  reconstruction error.} Suppose for instance that $m = 2n$. Then $s^2
= 4k(\log m)/m$ and we see that $s$ is small when there are few gross
errors. In this case, the recovery error is very close to that
attained by the ideal procedure.  Our experiments show that in
practice, the constant $C'_1$ is quite small: for instance, when $m =
2n$, one can correct 15\% of arbitrary errors, and in the overwhelming
majority of cases obtain a decoded vector whose MSE is less than 3
times larger than the ideal MSE.

Finally, this last result is in some way more subtle than the
corresponding result for the SOCP decoder.  Indeed,
\eqref{eq:Pinftydecode2} asserts that the accuracy of the LP decoder {\em
  automatically adapts to the number $k$ of gross errors} which were
introduced. The smaller this number, the smaller the recovery
error. For small values of $k$, the bound in \eqref{eq:Pinftydecode2} may
in fact be considerably smaller than its analog \eqref{eq:P2decode2}.

\section{Numerical Experiments}
\label{sec:simulations}

As mentioned earlier, numerical studies show that the empirical
performance of the proposed decoding strategies is noticeable.  To
confirm these findings, this section discusses an experimental setup
and presents numerical results. The reader wanting to reproduce our
results may find the matlab file available at
\texttt{http://www.acm.caltech.edu/$\sim$emmanuel/ConvexDecode.m}
useful. Here are the steps we used:
\begin{enumerate}
\item Choose a pair $(n,m)$ and sample an $m$ by $n$
  matrix $A$ with independent standard normal entries; the coding
  matrix is fixed throughout. 
\item Choose a fraction $\rho$ of grossly corrupted entries and define
  the number of corrupted entries as $k = \texttt{round}(\rho \cdot
  m)$; e.g. if $m = 512$ and 10\% of the entries are corrupted, $k = 51$.
\item Sample a block of information $x \in \R^n$ with independent
  and identically distributed Gaussian entries. Compute $A x$.
\item Select $k$ locations uniformly at random and flip the signs of
  $Ax$ at these locations.
\item Sample the vector $z = (z_1, \ldots, z_m)$ of smaller errors
  with $z_i$ i.i.d.~$N(0,\sigma^2)$, and add $z$ to the outcome of the
  previous step. Obtain $y$.
\item Obtain $\hat x$ by solving both $(P_2)$ and $(P_\infty)$ followed by
  a reprojection step discussed below \cite{Dantzig}.
\item Repeat steps (3)--(6) 500 times.  
\end{enumerate}

We briefly discuss the reprojection step. As observed in
\cite{Dantzig}, both programs $(P'_2)$ and $(P'_\infty)$ have a tendency to
underestimate the vector $e$ (they tend to be akin to
soft-thresholding procedures).  One can easily correct for this bias
as follows: 1) solve $(P'_2)$ or $(P'_\infty)$ and obtain $\hat e$; 2)
estimate the support of the gross errors $e$ via $I := \{i : |\hat
e_i| > \sigma\}$, where $\sigma$ is the standard deviation of the
smaller errors; recall that $y' := Q^* y = Q^* e + Q^* z$ and update the
estimate by regressing $y'$ onto the selected columns of $Q^*$ via the
method of least squares
\[
\hat e = \text{argmin } \|y'-Q^* \tilde e\|_{\ell_2}^2 \quad
\text{subject to} \quad \tilde{e}_i = 0, \, i \in I^c;
\]
3) finally, obtain $\hat x$ via $(A^* A)^{-1} A^*(y - \hat e)$ where
$\hat e$ is the reprojected estimate calculated in the previous
step. 

\newcommand{\rhoO}{\rho^{\text{Oracle}}}
\newcommand{\rhoI}{\rho^{\text{Ideal}}} 

In our series of experiments, we used $m = 2n = 512$ and a corruption
rate of 10\%. The standard deviation $\sigma$ is selected in such a
way that just about the first three binary digits of each entry of the
codeword $A x$ are reliable. Formally $\sigma =
\text{median}(Ax)/16$. Finally and to be complete, we set the
threshold $\eps$ in $(P_2)$ so that $\|Q^* z\|_{\ell_2} \le \eps$ with
probability .95; in other words, $\eps^2 = \chi^2_{m-n}(.95) \cdot
\sigma^2$, where $\chi^2_{m-n}(.95)$ is the 95th percentile of a
chi-squared distribution with $m-n$ degrees of freedom. We also set
the thresholds in the general formulation \eqref{eq:Pinftys} of $(P_\infty)$ in
a similar fashion. The distribution of $(QQ^* z)_i$ is normal with
mean 0 and variance $s_i^2 = (QQ^*)_{i,i} \cdot \sigma^2$ so that the
variable $z'_i = (QQ^* z)_i/s_i$ is standard normal. We choose
$\lambda_i = \lambda \cdot s_i$ where $\lambda$ obeys $$\sup_{1 \le i
  \le m} |z'_i| \le \lambda$$ with probability at least .95. In both
cases, our selection makes the true vector $e$ of gross errors 
feasible with probability at least .95. In our simulations, the
thresholds for the SOCP and LP decoders (the parameters
$\chi^2_{m-n}(.95)$ and $\lambda$) were computed by Monte Carlo
simulations.

To evaluate the accuracy of the decoders, we report two statistics
\begin{equation}
  \label{eq:ratios}
  \rhoI = \frac{\|\hat x - x\|}{\|x^{\text{Ideal}} - x\|}, \quad
\text{and} \quad \rhoO = \frac{\|\hat x - x\|}{\|x^{\text{Oracle}}-
  x\|},
\end{equation}
which compare the performance of our decoders with that of ideal
strategies which assume either exact knowledge of the gross errors or
exact knowledge of their locations. As discussed earlier,
$x^{\text{Ideal}}$ is the reconstructed vector one would obtain if the
gross errors were known to the receiver {\em exactly} (which is of
course equivalent to having no gross errors at all). The
reconstruction $x^{\text{Oracle}}$ is that one would obtain if,
instead, one had available an oracle supplying perfect information
about the location of the gross errors (but not their value).  Then
one could simply delete the corrupted entries of the received codeword
$y$ and reconstruct $x$ by the method of least squares, i.e.~find the
solution to $\|y^{\text{Oracle}} - A^{\text{Oracle}} \tilde
x\|_{\ell_2}$, where $A^{\text{Oracle}}$ (resp.~$y^{\text{Oracle}}$)
is obtained from $A$ (resp.~$y$) by deleting the corrupted rows.

The results are presented in Figure 1 and summarized in Table 1. These
results show that both our approaches work extremely well. As one can
see, our methods give reconstruction errors which are nearly as sharp
as if no gross errors had occurred or as if one knew the locations of
these large errors exactly. Put in a different way, the constants
appearing in our quantitative bounds are in practice very
small. Finally, the SOCP and LP decoders have about the same
performance although upon closer inspection, one could argue that the
LP decoder is perhaps a tiny bit more accurate.

\begin{figure}
\label{fig:histOracle}
\begin{tabular}{cc}
\includegraphics[width=2.75in]{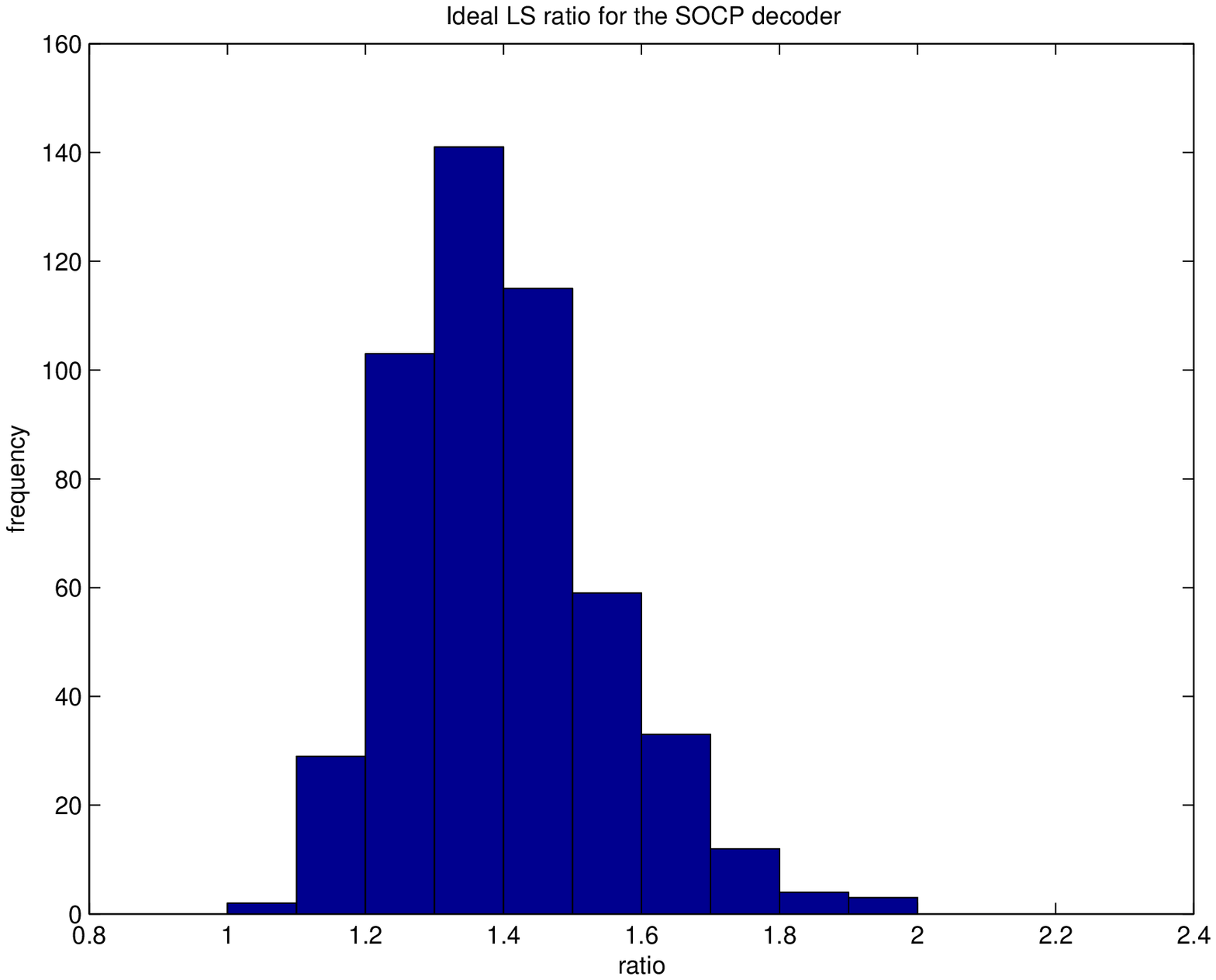} &
\includegraphics[width=2.75in]{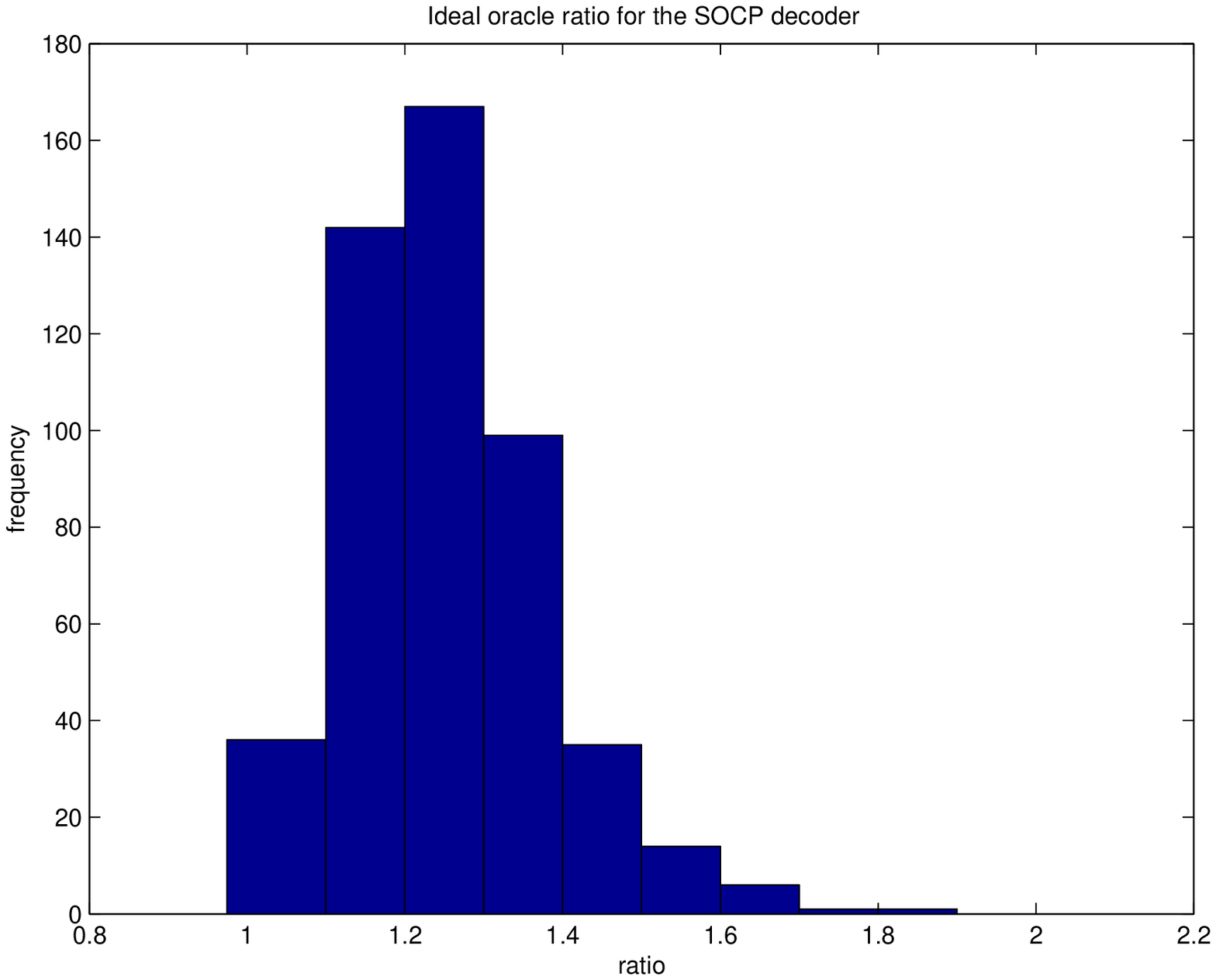}\\
\includegraphics[width=2.75in]{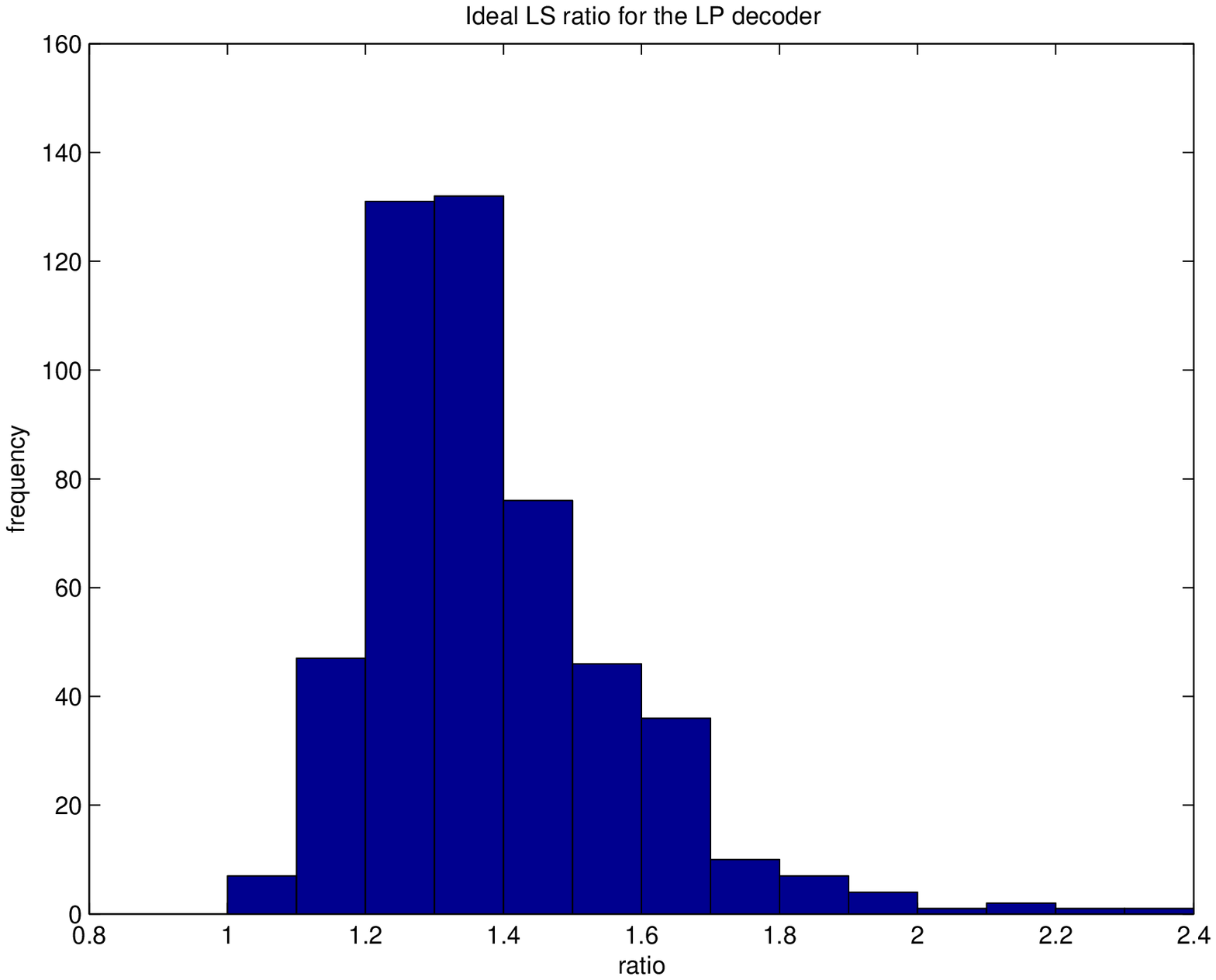} &
\includegraphics[width=2.75in]{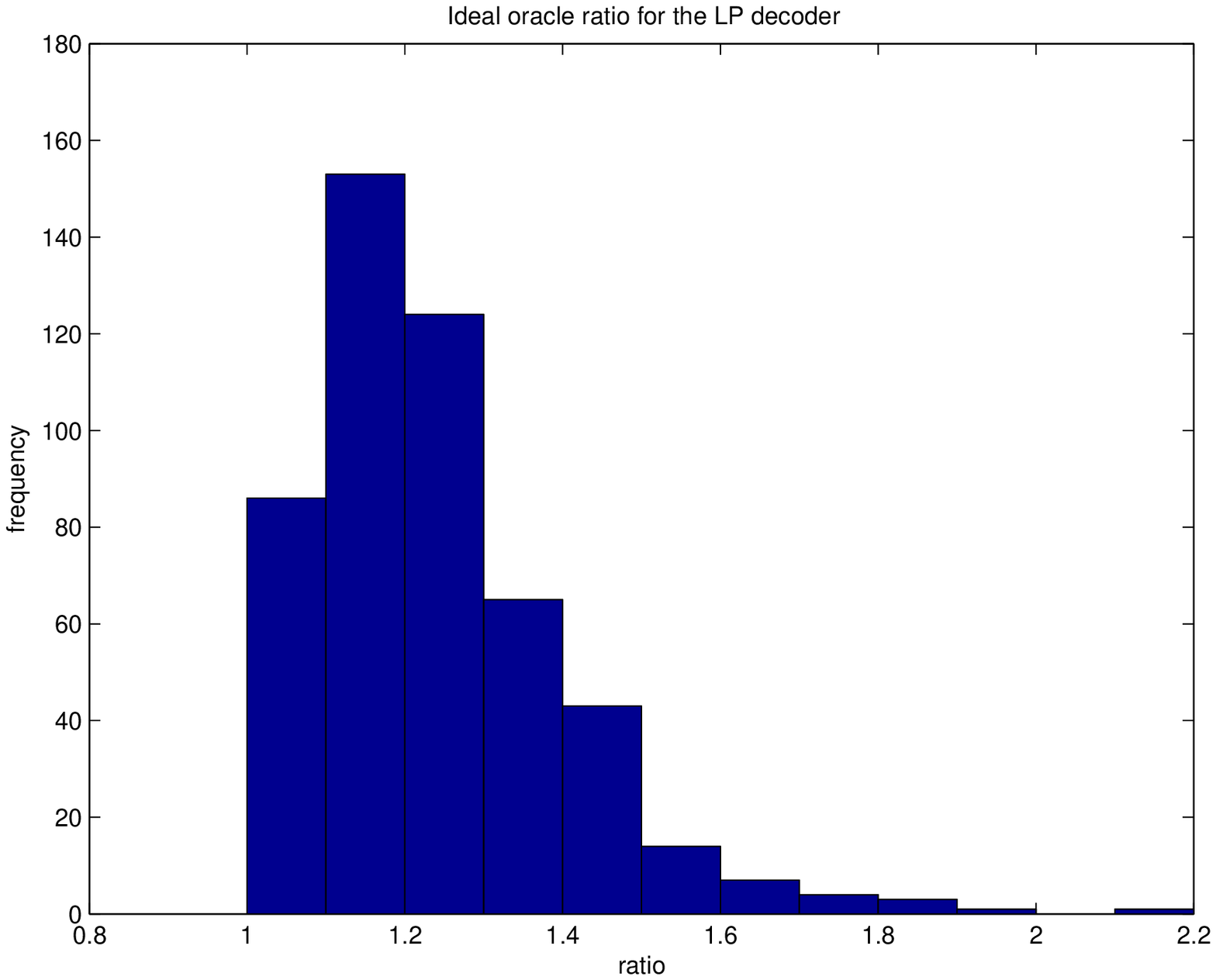}\\
\end{tabular}
\caption{Statistics of the ratios \eqref{eq:ratios} $\rhoI$ (first
  column) and $\rhoO$ (second column) which compare the performance of
  the proposed decoders with that of ideal strategies which assume
  either exact knowledge of the gross errors or exact knowledge of
  their locations. The first row shows the performance of the SOCP
  decoder, the second that of the LP decoder.}
\end{figure}

\begin{table}[htb]
  \begin{center}
        \begin{tabular}{|c|cc|cc|}
          \hline
          & median of $\rhoI$ & mean of $\rhoI$ & median of $\rhoO$  & 
          mean of $\rhoO$ \\
          \hline
          SOCP decoder & 1.386 & 1.401 & 1.241 & 1.253\\
          LP decoder & 1.346 & 1.386 & 1.212 & 1.239\\ 
          \hline
        \end{tabular}
  \end{center}  
  \caption{Summary statistics of the ratios $\rhoI$
    and $\rhoO$ \eqref{eq:ratios} for the Gaussian coding matrix.}
  \label{tbl:cmp}
\end{table}

We also repeated the same experiment but with a coding matrix $A$
consisting of $n = 256$ randomly sampled columns of the $512 \times
512$ discrete Fourier transform, and obtained very similar
results. The results are presented in Figure 2 and summarized in Table
2. The numbers are remarkably close to our earlier findings and again
both our methods work extremely well (again the LP decoder is a tiny
bit more accurate). This experiment is of special interest since it
suggests that one can apply our decoding algorithms to very large data
vectors, e.g. with sizes ranging in the hundred of thousands. The
reason is that one can use off-the-shelf interior point algorithms
which only need to be able to apply $A$ or $A^*$ to arbitrary vectors
(and never need to manipulate the entries of $A$ or even store
them). When $A$ is a partial Fourier transform, one can evaluate $A x$
and $A^* y$ by means of the FFT and, hence, this is well suited for
very large problems. See \cite{PracticalRecovery} for very large scale
experiments of a similar flavor.

\begin{figure}
\label{fig:histOracle2}
\begin{tabular}{cc}
\includegraphics[width=2.75in]{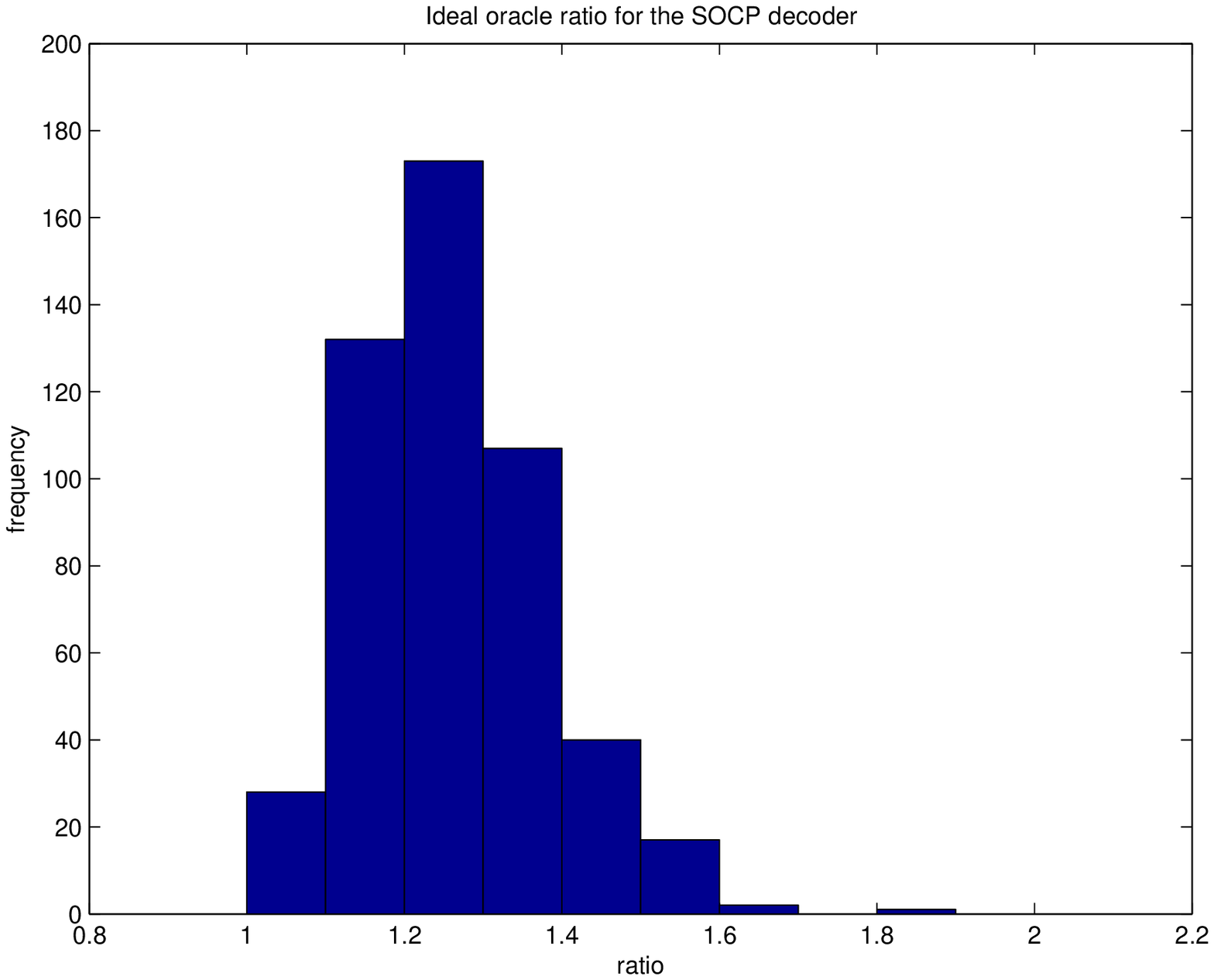} & 
\includegraphics[width=2.75in]{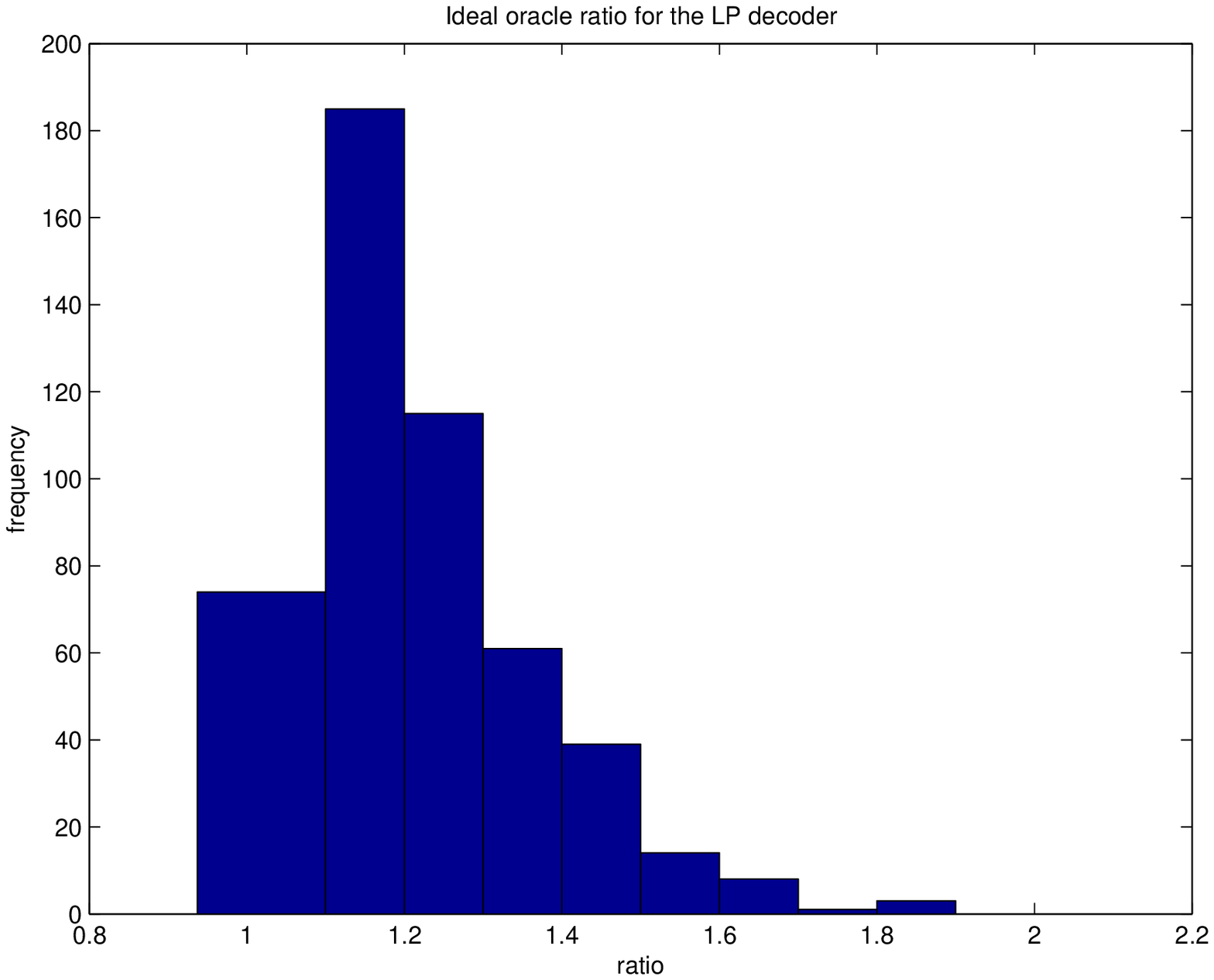}\\
\end{tabular}
\caption{Statistics of the ratios $\rhoO$ for the SOCP decoder (first
  column) and the LP decoder (second column) in the case where the
  coding matrix is a partial Fourier transform.}
\end{figure}

\begin{table}[htb]
  \begin{center}
        \begin{tabular}{|c|cc|cc|}
          \hline
          & median of $\rhoI$ & mean of $\rhoI$ & median of $\rhoO$  & 
          mean of $\rhoO$ \\
          \hline
          SOCP decoder &   1.390 &   1.401  &  1.244 &   1.262\\
          LP decoder &   1.337  &  1.375  &  1.195  &  1.230
          \\ 
          \hline
        \end{tabular}
  \end{center}  
  \caption{Summary statistics of the ratios $\rhoI$
    and $\rhoO$ \eqref{eq:ratios} for the Fourier coding matrix.}
  \label{tbl:cmp}
\end{table}

\section{Proofs}
\label{sec:proofs}

In this section, we prove all of our results. We begin with some
preliminaries which will be used throughout, then prove the claims
about the SOCP decoder, and end this section with the LP decoder.  Our
work builds on \cite{CRT2} and \cite{Dantzig}.

\subsection{Preliminaries}
\label{sec:preliminaries}

We shall make extensive use of two simple lemmas that we now record.
\begin{lemma}
  Let $Y_d \sim \chi^2_d$ be distributed as a chi-squared random
  variable with $d$ degrees of freedom. Then for each $t > 0$  
  \begin{equation}
    \label{eq:chi2}
    \P(Y_d - d \ge t \, \sqrt{2d}  + t^2)  
\le e^{-t^2/2} \quad 
\text{ and } \quad  \P(Y_d - d \le -t \, 
\sqrt{2d}) \le e^{-t^2/2}.  
  \end{equation}
\end{lemma}

This is fairly standard \cite{LaurentMassart}, see also
\cite{chi2oracle} for very slightly refined estimates. A consequence
of these large deviation bounds is the estimate below.

\begin{lemma}
\label{teo:norms}
Let $U = (u_1, u_2, \ldots, u_m)$ be a vector uniformly distributed on
the unit sphere in $m$ dimensions. Let $Z_n = u_1^2 + \ldots + u_n^2$
be the squared length of the first $n$ components of $u$. Then for
each $t \le 1/2$
\begin{equation}
\label{eq:norms}
\P\left(Z_n \le \frac{n}{m} \, (1- t) \right) \le e^{-nt^2/16} + e^{-mt^2/24}. 
\end{equation}
\end{lemma}
\begin{proof}
  Suppose $X_1, X_2, \ldots, X_m$ are i.i.d.~$N(0,1)$. Then the
  distribution of $U$ is that of the vector $X/\|X\|_{\ell_2}$ and,
  therefore, the law of $Z_n$ is that of $Y_n/Y_m$, where $Y_k =
  \sum_{j \le k} X_j^2$.  Define the events $A_\eps = \{Y_m \ge
  (1+\eps)m\}$ for some $\eps > 0$ and $B_t = \{Y_n/Y_m \le
  n/m(1-t)\}$.  We have
  \begin{align*}
    \P(B_t) & = \P(B_t \, | \,
    A_\eps^c) \, \P(A_\eps^c) + \P(B_t \, | \, 
    A_\eps) \, \P(A_\eps)\\
    & \le \P(Y_n \le  n(1-t)(1+\eps)) + \P(Y_m \ge (1+\eps) m).
  \end{align*}
With $\eps = t/2$, this gives  $n(1-t)(1+\eps) \le n(1-t/2)$ and thus  
 \begin{align*}
   \P(Z_n \le n/m(1-t)) & \le \P(Y_n \le
   n(1-t/2)) + \P(Y_m \ge (1+\eps)m)\\
   & \le e^{-nt^2/16} + e^{-\gamma^2 m/2},
 \end{align*}
 which follows from \eqref{eq:chi2}, where $\gamma$ obeys $t/2 =
 \gamma \sqrt{2} + \gamma^2$. For $t \le 1/2$, $\gamma \le
 1/2\sqrt{3}$ (for small values of $t$, $\gamma \approx t/2\sqrt{2}$)
 and the conclusion follows. 
\end{proof}

\subsection{The SOCP decoder}
\label{sec:proofsocp}

For a matrix $\Phi$, define the sequences $(a_k)$ and $(b_k)$ as
respectively the largest and smallest numbers obeying
\begin{equation}
  \label{eq:iso-ext}
  a_k \|x\|_{\ell_2} \le \|\Phi x\|_{\ell_2} \le b_k \|x\|_{\ell_2}, 
\end{equation}
for all $k$-sparse vectors. In other words, if we list all the
singular values of all the submatrices of $\Phi$ with $k$ columns,
$a_k$ is the smallest element from that list and $b_k$ the largest.
Note of course the resemblance with \eqref{eq:rip}---only this is
slightly more general. We now adapt an important result from
\cite{CRT2}.
\begin{lemma}[adapted from \cite{CRT2}]
  \label{teo:CRT2}
  Set $\Phi \in \R^{r \times m}$ and let $(a_k)$ and $(b_k)$ be the
  restricted extremal singular values of $\Phi$ as in
  \eqref{eq:iso-ext}.  Any point $\tilde x \in \R^m$ obeying 
\begin{equation}
\label{eq:conetube}
\|\tilde x\|_{\ell_1} \le \|x\|_{\ell_1}, \quad 
\text{ and } \quad \|\Phi \tilde x - \Phi x\|_{\ell_2} \le 2\eps, 
\end{equation}
also obeys 
  \begin{equation}
    \label{eq:CRT2}    
\|\tilde x - x\|_{\ell_2} \le 
\frac{\sqrt{6} \, \eps}{a_{3k}(\Phi) - \frac{1}{\sqrt{2}}\, b_{2k}(\Phi)},  
  \end{equation}
  provided that $x$ is $k$-sparse with $k$ such that $a_{3k}(\Phi) -
  \frac{1}{\sqrt{2}}\, b_{2k}(\Phi) > 0$. 
\end{lemma}
The proof follows the same steps as that of Theorem 1.1 in
\cite{CRT2}, and is omitted. In particular, it follows from (2.6) in
the aforementioned reference with $M = 2 |T_0|$ and $a_{M + |T_0|}$
(resp.~$b_M$) in place of $\sqrt{1-\delta_{|T_0| + M}}$
(resp.~$\sqrt{1+\delta_{M}})$ in the definition of $C_{|T_0|, M}$.

\subsubsection{Proof of Theorem \ref{teo:P2decode}}  
Recall that the solution $(\hat x, \hat z)$ to $(P_2)$ obeys
\eqref{eq:correctLS} where $\hat e$ is the solution to
$(P'_2)$. Replacing $y$ in \eqref{eq:correctLS} with $A x + e + z$
gives
\begin{align}
  \nonumber
  \hat x - x & = (A^* A)^{-1} A^* (e-\hat e) +   (A^* A)^{-1} A^* z\\
  \label{eq:easy} 
& = (A^* A)^{-1} A^* (e-\hat e) +   x^{\text{Ideal}} - x, 
\end{align}
and since $A^* A = I$, 
\[
\|\hat x - x \|_{\ell_2} \le \|A^*(e-\hat e)\|_{\ell_2} + \|x^{\text{Ideal}} -
x\|_{\ell_2}.
\]
To prove \eqref{eq:P2decode}, it then suffices to show that $\|e -
\hat e\|_{\ell_2} \le \frac{C \, \eps}{\sqrt{1-\frac{n}{m}}}$ since
the 2-norm of $A^*$ is at most 1.

By assumption $\|Q^*(y-e)\|_{\ell_2} = \|Q^*z\|_{\ell_2} \le \eps$ and
thus, $e$ is feasible for $(P'_2)$ which implies $\|\hat e\|_{\ell_1}
\le \|e\|_{\ell_1}$. Moreover, $$\|Q^*e - Q^*\hat e\|_{\ell_2} \le
\|Q^*(y-e)\|_{\ell_2} + \|Q^*(y-\hat e)\|_{\ell_2} \le 2\eps.$$ We
then apply Lemma \ref{teo:CRT2} (with $\Phi = Q^*$) and obtain
\begin{equation}
  \label{eq:intermediatea}
  \|e - \hat e\|_{\ell_2} \le \frac{\sqrt{6} \, \eps}{a_{3k}(Q^*) -
  \frac{1}{\sqrt{2}}\, b_{2k}(Q^*)}.
\end{equation}
Now since the $m \times m$ matrix obtained by concatenating the
columns of $A$ and $Q$ is an isometry, we have
\[
\|A^* x\|_{\ell_2}^2 + \|Q^* x\|_{\ell_2}^2 = \|x\|_{\ell_2}^2 \quad \forall x
\in \R^m, 
\]
whence
\begin{align*}
  a^2_{k}(Q^*) & = 1  - b^2_k(A^*),\\
  b^2_{k}(Q^*) & = 1 - a^2_k(A^*).
\end{align*}
Assuming that $a_{3k}(Q^*) \ge \frac{1}{\sqrt{2}} b_{2k}(Q^*)$, 
we deduce from \eqref{eq:intermediatea} that
\begin{equation}
\label{eq:intermediateb} 
\|e - \hat e\|_{\ell_2} \le \sqrt{6} \, \eps \cdot \frac{a_{3k}(Q^*) +
  \frac{1}{\sqrt{2}}  b_{2k}(Q^*)}{1-b^2_{3k}(A^*) -
  \frac{1}{2} (1-a^2_{2k}(A^*))} \le 2\sqrt{6} \eps \cdot
\frac{a_{3k}(Q^*)}{\frac{1}{2} + \frac{1}{2}a^2_{2k}(A^*) -
  b_{3k}^2(A^*)}.
\end{equation}
Recall that $(\delta_k)$ are the restricted isometry constants of
$\sqrt{\frac{m}{n}} \, A^*$, and observe that by definition for each
$k = 1, 2, \ldots$,
\[
a_k^2(A^*) \ge \frac{n}{m} (1-\delta_k), \qquad b^2_k(A^*) \le
\frac{n}{m} (1+\delta_k). 
\]
It follows that the denominator on the right-hand side of
\eqref{eq:intermediateb} is greater or equal to 
$$ 
\frac{1}{2} + \frac{n}{2m}
(1-\delta_{2k}) - \frac{n}{m}(1+\delta_{3k}) = 
\frac{1}{2}\left(1-\frac{n}{m}\right) - \frac{n}{m}\left(\delta_{3k} +
  \frac{1}{2}\delta_{2k}\right).
$$
Now suppose that for some $0 < c < 1$,
\[
\delta_{3k} + \frac{1}{2}\delta_{2k} \le \frac{c}{2} \cdot
\left(\frac{m}{n} - 1\right). 
\]
This automatically implies $a_{3k}(Q^*) \ge \frac{1}{\sqrt{2}}
b_{2k}(Q^*)$, and the denominator on the right-hand side of
\eqref{eq:intermediateb} is greater or equal to
$\frac{1}{2}(1-c)(1-\frac{n}{m})$. The numerator obeys 
\[
a^2_{3k}(Q^*) = 1 - b^2_{3k}(A^*) \le 1 - a^2_{3k}(A^*) \le 1 -
(1-\delta_{3k}) \frac{n}{m}.
\]
Since $\frac{n}{m} \, \delta_{3k} \le \frac{c}{2}(1-\frac{n}{m})$, we
also have $a^2_{3k}(Q^*) \le (1 + \frac{c}{2})(1-\frac{n}{m})$.  In
summary, \eqref{eq:intermediateb} gives
\[
\|e - \hat e\|_{\ell_2} \le C_2 \cdot
\frac{\eps}{\sqrt{1-\frac{n}{m}}},
\]
where one can take $C_2$ as $4\sqrt{6(1+c/2)}/(1-c)$.  This
establishes the first part of the claim.

We now turn to the second part of the theorem and argue that if the
orthonormal columns of $A$ are chosen uniformly at random, the error
bound \eqref{eq:P2decode} is valid as long as we have a constant
fraction of gross errors. Put $r = m-n$ and let $X$ be an $m$ by $r$
matrix with independent Gaussian entries with mean 0 and variance
$1/m$. Consider now the reduced singular value decomposition of $X$
\[
X = U \Sigma V^*, \quad U \in \R^{m \times r} \text{ and } \Sigma, V \in \R^{r
  \times r}.
\]
Then the columns of $U$ are $r$ orthonormal vectors selected uniformly
at random and thus $U$ and $Q$ have the same distribution. Thus we
can think of $Q$ as being the left singular vectors of a Gaussian
matrix $X$ with independent entries. From now on, we identify $U$ with
$Q$. Observe now that
\[
\|X^*(\hat e - e)\|_{\ell_2} = \|V\Sigma Q^*(\hat e -e)\|_{\ell_2} = \|\Sigma
Q^*(\hat e -e)\|_{\ell_2} \le \sigma_1(X) \, \|Q^*(\hat e -e)\|_{\ell_2}, 
\]
where $\sigma_1(X)$ is the largest singular value of $X$. The singular
values of Gaussian matrices are well concentrated and a classical
result \cite{Szarek1} shows that 
\begin{equation}
\label{eq:ledoux1}
\P\left(\sigma_{1}(X) > 1 + \sqrt{\frac{r}{m}} + t \right)  \le e^{-m t^2/2}.
\end{equation}
By choosing $t = 1$ in the above formula, we have 
\[
\|X^*(\hat e - e)\|_{\ell_2} \le 3 \|Q^*(\hat e -e)\|_{\ell_2} \le 6\eps
\]
with probability at least $1 - e^{-m/2}$ since $\|Q^*(\hat e
-e)\|_{\ell_2} \le 2\epsilon$. We now apply Lemma \ref{teo:CRT2}
with $\Phi = X^*$, which gives
\begin{equation}
  \label{eq:intermediate}
  \|e - \hat e\|_{\ell_2} \le \frac{3\sqrt{6} \, \eps}{a_{3k}(X^*) -
    \frac{1}{\sqrt{2}}\, b_{2k}(X^*)} =  \sqrt{\frac{m}{r}} \cdot   
  \frac{3\sqrt{6} \, \eps}{a_{3k}(Y^*) - 
    \frac{1}{\sqrt{2}}\, b_{2k}(Y^*)}, 
\end{equation}
where $Y = \sqrt{\frac{m}{r}} X$. The theorem is proved since it is
well known that if $k \le c_0 \cdot r/\log(m/r)$ for some constant
$c_0$, we have $a_{3k}(Y^*) - \frac{1}{\sqrt{2}}\, b_{2k}(Y^*) \ge
c_1$ with probability at least $1 - O(e^{-\gamma' r})$ for some
universal constants $c_1$ and $\gamma$; this follows from available
bounds on the restricted isometry constants of Gaussian matrices
\cite{OptimalRecovery,DecodingLP,Equivl0l1,rudelson06sp}.

\subsubsection{Proof of Corollary \ref{cor:P2decode}} First, we can
just assume that $\sigma = 1$ as the general case is treated by a
simple rescaling.  Put $r = m-n$. Since the random vector $z$ follows
a multivariate normal distribution with mean zero and covariance
matrix $I_m$ ($I_m$ is the identity matrix in $m$ dimensions), $Q^* z$
is also multivariate normal with mean zero and covariance matrix $Q^*Q
= I_r$. Consequently, $\|Q^* z\|_{\ell_2}^2$ is distributed as a
chi-squared variable with $r$ degrees of freedom. Pick $\lambda =
\gamma \, \sqrt{r}$ in \eqref{eq:chi2}, and obtain
\[
\P\left(\|Q^*z\|^2_{\ell_2} \ge (1+ \gamma \sqrt{2} +\gamma^2)r \right) \le e^{-\gamma^2 r/2}.
\]
With $t = \gamma \sqrt{2} +\gamma^2$ so that $\gamma =
(\sqrt{1+2t}-1)/\sqrt{2}$, we have $\|Q^*z\|_{\ell_2}\le
\sqrt{r(1+t)}$ with probability at least $1-e^{-\gamma^2(m-n)/2}$. On
this event, Theorem \ref{teo:P2decode} asserts that
\[
\|\hat x - x\|_{\ell_2} \le C \, \sqrt{m(1+t)} + \|x -
x^{\text{Ideal}}\|_{\ell_2}. 
\]
This essentially concludes the proof of the corollary since the size
of $\|x - x^{\text{Ideal}}\|_{\ell_2}$ is about $\sqrt{n}$.  Indeed,
$\|x - x^{\text{Ideal}}\|^2_{\ell_2} = \|A^* z\|^2_{\ell_2} \sim
\chi^2_n$ as observed earlier. As a consequence, for each $t_0 > 0$, we
have $\|x - x^{\text{Ideal}}\|_{\ell_2} \le \sqrt{n(1+t_0)} \cdot
\sigma$ with probability at least $1-e^{-\gamma_0^2 n/2}$, where
$\gamma_0$ is the same function of $t_0$ as before. Selecting $t_0$ as
$t_0 = m/n$, say, gives the result.

\subsection{The LP decoder}
\label{sec:prooflp}

Before we begin, we introduce the number $\theta_{k,k'}$ of a matrix
$\Phi \in \R^{r \times m}$ for $k + k' \le m$ called the
\emph{$k,k'$-restricted orthogonality constants}. This is the smallest
quantity such that
\begin{equation}\label{ortho}
 |\langle \Phi v, \Phi v' \rangle|  \leq \theta_{k,k'} \cdot  
\|v\|_{\ell_2} \, \|v'\|_{\ell_2} 
 \end{equation}
 holds for all $k$ and $k'$-sparse vectors supported on disjoint sets.
 Small values of restricted orthogonality constants indicate that
 disjoint subsets of columns span nearly orthogonal subspaces. The
 following lemma which relates the number $\theta_{k,k'}$ to the
 extremal singular values will prove useful.
\begin{lemma}
\label{teo:abtheta}
For any matrix $\Phi \in \R^{r \times m}$, we have 
\[
\theta_{k,k'}(\Phi) \le \frac{1}{2} \, (b_{k+k'}^2(\Phi) - a^2_{k+k'}(\Phi)).
\] 
 \end{lemma}
\begin{proof}
  Consider two vectors $v$ and $v'$ which are respectively $k$ and
  $k'$-sparse. By definition we have 
  \begin{align*}
    2 a^2_{k+k'}(\Phi) & \le \|\Phi v + \Phi v'\|_{\ell_2}^2  \le 2
    b^2_{k+k'}(\Phi),\\
  2 a_{k+k'}^2(\Phi) & \le \|\Phi v - \Phi v'\|_{\ell_2}^2 \le 2
    b^2_{k+k'}(\Phi),
  \end{align*}
and the conclusion follows from the parallelogram identity 
\[
|\<\Phi v, \Phi v'\>| = \frac{1}{4} \left| \|\Phi v + \Phi
  v'\|_{\ell_2}^2 - \|\Phi v - \Phi v'\|_{\ell_2}^2\right| \le
\frac{1}{2} \, (b_{k+k'}^2(\Phi) - a^2_{k+k'}(\Phi)).
\]
\end{proof}

The argument underlying Theorem \ref{teo:Pinftydecode} uses an
intermediate result whose proof may be found in the Appendix. Here and
in the remainder of this paper, $x_I$ is the restriction of the vector
$x$ to an index set $I$, and for a matrix $X$, $X_I$ is the submatrix
formed by selecting the columns of $X$ with indices in $I$.
\begin{lemma}
  \label{teo:DS}
  Let $\Phi$ be an $r\times m$-dimensional matrix and suppose $T_0$ is
  a set of cardinality $k$.  For a vector $h \in \R^m$, we let $T_1$
  be the $k'$ largest positions of $h$ outside of $T_0$. Put $T_{01} =
  T_0\cup T_1$. Then
\begin{equation}
\label{eq:usefulDS}
\|h_{T_{01}}\|_{\ell_2} \leq \frac{1}{a^2_{k+k'}(\Phi)} \, 
\|\Phi^*_{T_{01}} \Phi h\|_{\ell_2} + \frac{\theta_{k',k+k'}(\Phi)}
{a^2_{k+k'}(\Phi) \, \sqrt{k'}} \, \|h_{T_0^c}\|_{\ell_1}
\end{equation}
and
\begin{equation}
\label{eq:usefulDS2}
\|h\|^2_{\ell_2} \leq \|h_{T_{01}}\|^2_{\ell_2} + \frac{1}{k'} 
\| h_{T_0^c}\|_{\ell_1}^2.
\end{equation}
\end{lemma}

\subsubsection{Proof of Theorem \ref{teo:Pinftydecode}} Just as before, it
suffices to show that $\|e - \hat e\|_{\ell_2} \le C \sqrt{k} \cdot
\lambda \cdot (1-n/m)^{-1}$.  Set $h = \hat e - e$ and let $T_0$ be
the support of $e$ (which has size $k$). Because $e$ is feasible for
$(P'_\infty)$ we have on the one hand $\|\hat e\|_{\ell_1} \le
\|e\|_{\ell_1}$, which gives
\[
\|e_{T_0}\|_{\ell_1} - \|h_{T_0}\|_{\ell_1} + \|h_{T_0^c}\|_{\ell_1}
\le \|e+h\|_{\ell_1} \le \|e\|_{\ell_1} \Rightarrow
\|h_{T_0^c}\|_{\ell_1} \le \|h_{T_0}\|_{\ell_1}. 
\]
Note that this has an interesting consequence since 
\begin{equation}
  \label{eq:l1l2}
  \|h_{T_0^c}\|_{\ell_1}  \le  \|h_{T_0}\|_{\ell_1} \le \sqrt{k} \cdot 
\|h_{T_0}\|_{\ell_2}
\end{equation}
by Cauchy Schwarz. On the other hand 
\begin{equation}
  \label{eq:tube}
  \|QQ^* h\|_{\ell_\infty} \le \|QQ^* (\hat e - y)\|_{\ell_\infty} + 
\|QQ^* (y-e)\|_{\ell_\infty} \le 2\lambda. 
\end{equation}

The ingredients are now in place to establish the claim. We set $k' =
k$, apply Lemma \eqref{teo:DS} with $\Phi = Q^*$ to the vector $h =
\hat e - e$, and obtain
\begin{equation}
  \label{eq:almost1}
  \|h\|_{\ell_2} \le \sqrt{2} \, \|h_{T_{01}}\|_{\ell_2}, 
  \quad \text{ and } \quad \|h_{T_{01}}\|_{\ell_2} \le  
\frac{1}{a^2_{2k}(Q^*) - \theta_{k,2k}(Q^*)} \, \|Q_{T_{01}} Q^*h\|_{\ell_2}. 
\end{equation}
Since each component of $Q_{T_{01}} Q^*h$ is at most equal to
$2\lambda$, see \eqref{eq:tube}, we have $\|Q_{T_{01}} Q^*h\|_{\ell_2}
\le \sqrt{2k} \cdot 2\lambda$.  We then conclude from Lemma
\ref{teo:abtheta} that
\begin{equation}
  \label{eq:almost2}
  \|h\|_{\ell_2} \le 2\sqrt{k} \cdot   
\frac{2\lambda}{a^2_{2k}(Q^*) + 
\frac{1}{2}a^2_{3k}(Q^*) - \frac{1}{2}b^2_{3k}(Q^*)}. 
\end{equation}
For each $k$, recall the relations $a_{k}^2(Q^*) = 1 - b_{k}^2(A^*)$
and $b_{k}^2(Q^*) = 1 - a_{k}^2(A^*)$ which give
\[
\|h\|_{\ell_2} \le 4\sqrt{k} \cdot \frac{\lambda}{D}, \qquad D:=
1-b^2_{2k}(A^*) - \frac{1}{2}b^2_{3k}(A^*) + \frac{1}{2}a^2_{3k}(A^*).
\]
Now just as before, it follows from our definitions that for each $k$,
$b^2_{k}(A^*) \le \frac{n}{m}(1+\delta_k)$ and $a^2_{k}(A^*) \ge
\frac{n}{m}(1-\delta_k)$. These inequalities imply 
\[
D \ge 1-\frac{n}{m}(1+\delta_{2k} + \delta_{3k}). 
\]
Therefore, if one assumes that 
\[
\delta_{2k} + \delta_{3k} \le c\left(\frac{m}{n}-1\right),
\]
for some fixed constant $0 < c < 1$, then 
\[
\|e-\hat e\|_{\ell_2} = \|h\|_{\ell_2} \le \frac{4\sqrt{k}}{1-c} \cdot
\frac{\lambda}{1-\frac{n}{m}}.
\]
This establishes the first part of the theorem.

We turn to the second part of the claim; if the orthonormal columns of
$A$ are chosen uniformly at random, we show that the error bound
\eqref{eq:Pinftydecode} is valid with large probability as long as we have
a constant fraction of gross errors. The same argument as before
(albeit for a general value of $k'$) gives
\begin{equation}
\label{eq:almost3}
\|h\|_{\ell_2} \le \frac{k+k'}{\sqrt{k'}} \cdot
\frac{2\lambda}{D}, \qquad D := a^2_{k+k'}(Q^*) + 
\frac{1}{2}\sqrt{\frac{k}{k'}}(a^2_{k+2k'}(Q^*) -
b^2_{k+2k'}(Q^*)).
\end{equation}
so that this is really a question about the extremal singular values
of random orthogonal projections when restricted to sparse inputs.

Put $r = m-n$ and let $X$ be an $m$ by $r$ matrix with independent
Gaussian entries with mean 0 and variance $1/m$. Recall the QR
factorization of $X$.
\[
X = Q' R, \quad Q'\in \R^{m\times r}, \, R \in \R^{r\times r}, 
\]
where $R$ is upper triangular. The columns of $Q'$ are $r$ orthonormal
vectors selected uniformly at random and thus, $Q$ and $Q'$ have the
same distribution so that we can think of $Q$ as being the $Q$-factor
in the QR factorization of $X$. Also observe that $\sigma_j(X) =
\sigma_j(R)$, $1 \le j \le r$, i.e.~the nonzero singular values of $R$
and $X$ coincide. It follows from
\[
v^* X X^* v = v^* QRR^*Q^* v
\]
(which is valid for all $v \in R^m$) that 
\[
\sigma_r(X) \|Q^*v\|_{\ell_2} \le \|X^* v\|_{\ell_2} \le \sigma_1(X)
\|Q^*v\|_{\ell_1}. 
\]
Applying the above inequalities to $k$-sparse vectors gives 
\begin{align*}
b_k(Q^*) & \le \frac{1}{\sigma_r(X)} b_k(X^*),\\
a_k(Q^*) & \ge \frac{1}{\sigma_1(X)} a_k(X^*). 
\end{align*}
The point is that the extremal singular values $(a_k(X^*),b_k(X^*))_{1
  \le k \le r}$ are perhaps easier to study than those of $Q^*$.

Indeed, classical results from random matrix theory
\cite{Szarek1,LedouxAMS} assert that for each $t > 0$, 
\begin{align}
  \label{eq:ledoux2a}
  \P\left(\sigma_{1}(X^*) > 1 + (1+t)
    \sqrt{\frac{r}{m}} \right) & \le
  e^{-r t^2/2},\\
\label{eq:ledoux2b}
  \P\left(\sigma_{r}(X^*) <
    1 - (1+t)\sqrt{\frac{r}{m}}\right) & \le e^{-r t^2/2}.
\end{align}
These inequalities can be specialized to $r \times k$ submatrices of
$X$, and taking the union bound also show that for each $t > 0$
\begin{align}
\label{eq:uup1}
  \P\left(\sqrt{\frac{m}{r}} \, b_k(X^*)  > 1 + 
    \sqrt{\frac{k}{r}} + t \right) & \le \binom{m}{k} \, 
  e^{-r t^2/2},\\
\label{eq:uup2}
\P \left(\sqrt{\frac{m}{r}} \, a_k(X^*)  < 1 - 
    \sqrt{\frac{k}{r}} - t \right) & \le \binom{m}{k} \, 
  e^{-r t^2/2}.
\end{align}

We use these estimates to bound below the denominator $D$ in
\eqref{eq:almost3}. We first study the case where $r \le m/16$ and in
the sequel, we will denote $\sqrt{k/k'}$ by $\rho$. First, pick $t =
1/3$ in \eqref{eq:ledoux2a}--\eqref{eq:ledoux2b}. Then the event
\[
E_0 := \{2/3 \le \sigma_r(X) \le \sigma_1(X) \le 4/3\}
\]
has probability at least $1-2e^{-r/18}$.  On this event, the
denominator $D$ in \eqref{eq:almost3} obeys
\begin{equation*}
  D  \ge (3/4)^2 a^2_{k+k'}(X^*) - (3/2)^2 \frac{\rho}{2} b^2_{k+2k'}(X^*) = 
 \frac{9}{16} \, (a^2_{k+k'}(X^*) - 2\rho \, b^2_{k+2k'}(X^*)). 
\end{equation*}
Second, selecting $t = 1/8$ in \eqref{eq:uup1} and \eqref{eq:uup2}
shows that the events $E_1$ and $E_2$ respectively equal to 
\[
\sqrt{\frac{m}{r}} a_{k+k'}(X^*) \ge \frac{7}{8} -
  \sqrt{\frac{k+k'}{r}} \quad \text{ and } \quad 
\sqrt{\frac{m}{r}} b_{k+2k'}(X^*) > \frac{9}{8} +
  \sqrt{\frac{k+2k'}{r}} 
\]
have probability at least $1- \binom{m}{k+k'} e^{-r/128}$ and $1 -
\binom{m}{k+2k'} e^{-r/128}$.  Third, select $k'$ to be the smallest
integer so that $\rho = \sqrt{k/k'} \le 1/8$. Combining these facts
gives 
\[
\P\left(D \ge \frac{9r}{16m} \, \left[\left(\frac{7}{8} -
      \sqrt{\frac{k+k'}{r}}\right)^2 - \frac{1}{4} \left(\frac{9}{8} +
      \sqrt{\frac{k+2k'}{r}}\right)^2 \right]\right) \ge 1-\eta,
\]
where $\eta = 2\binom{m}{k+2k'} e^{-r/128} + 2 e^{-r/18}$. Elementary
calculations show that 
\[
\left(\frac{7}{8} - \sqrt{\frac{k+k'}{r}}\right) - \frac{1}{2}
\left(\frac{9}{8} + \sqrt{\frac{k+2k'}{r}}\right) \ge \frac{1}{16} \quad
\text{if} \quad \sqrt{\frac{k+k'}{r}} +
\frac{1}{2}\sqrt{\frac{k+2k'}{r}} \le \frac{1}{4},
\]
and 
\[
\left(\frac{7}{8} - \sqrt{\frac{k+k'}{r}}\right) - \frac{1}{2}
\left(\frac{9}{8} + \sqrt{\frac{k+2k'}{r}}\right) \ge 1
\]
under the same condition. In summary, $D \ge {9r}/{256m}$ with
probability at least $1-\eta$ provided that
\[
\sqrt{\frac{k+k'}{r}} + \frac{1}{2}\sqrt{\frac{k+2k'}{r}} \le
\sqrt{\frac{k}{r}} \cdot \left(\sqrt{66} + \frac{1}{2}
  \sqrt{131}\right) \le \frac{1}{4}
\] 
since $k'/k \le 65$. Finally $k + 2k' \le 131 k$ and assuming $131 k
\le m/2$, we also have
\begin{equation}
\label{eq:six}
\log \binom{m}{k+2k'} \le \log \binom{m}{131 k} \le 131 k
\left[1+ \log \frac{m}{131 k}\right].
\end{equation}
In other words, $D \ge {9r}/{256 m}$ with probability at least
$1-2e^{-r/256} - 2e^{-r/18}$ as long as the right-hand side of
\eqref{eq:six} is less or equal to $r/256$.  It follows from
$\eqref{eq:almost3}$ that with at least the same probability, $\|\hat
e - e\|_{\ell_2} \le C_1 \cdot (m/r) \cdot \sqrt{k} \cdot \lambda$ and
hence, one can correct a constant fraction of errors (the fraction
depends on $n/m$ of course) in the case where $r \le m/16$.

It remains to argue that the result is valid when $r \ge m/16$.  Here,
 the denominator $D$ in
\eqref{eq:almost3} obeys 
\[
D \ge b_{k+2k'}^2(Q^*) -\rho/2 \ge  b_{1}^2(Q^*) - \rho/2. 
\]
Let $Z_r$ be the squared $\ell_2$ norm of the first column of $Q^*$,
i.e.~$Z_r = (QQ^*)_{1,1}$. We have $b_1^2(Q^*) \ge Z_r$ and moreover,
Lemma \ref{teo:norms} proved that
\[
\P(Z_r < 3r/4m) \le 2e^{-\gamma_0^2 r/2}
\]
for some constant $\gamma_0 > 0$. Pick $k'$ to be the smallest integer
so that $\rho = \sqrt{k/k'} \le r/m$.  Then $D \ge r/4m$, and we
conclude that $\|\hat e - e\|_{\ell_2} \le C \cdot (m/r) \cdot
\sqrt{k} \cdot \lambda$ with probability at least $1 - 2e^{-\gamma_0^2
  r/2}$. To be complete, the condition on $k$ is $k + 2k' \le r$
which is satisfied if $k(3+(m/r)^2) \le r$ or equivalently $k \le
\rho_0 \cdot m$ with $\rho_0 = r/m \cdot (1+(m/r)^2)^{-1}$.

\subsubsection{Proof of Corollary \ref{cor:Pinftydecode}} First, we can
just assume that $\sigma = 1$ as the general case is treated by a
simple rescaling.  The random vector $QQ^* z$ follows a multivariate
normal distribution with mean zero and covariance matrix $QQ^*$. In
particular $(QQ^* z)_i \sim N(0,s^2_i)$, where $s_i^2 =
(QQ^*)_{i,i}$. This implies that $z'_i = (QQ^* z)_i/s_i$ is standard
normal with density $\phi(t) = (2\pi)^{-1/2} e^{-t^2/2}$. For each
$i$, $\P(|z'_i| > t) \le \phi(t)/t$ and thus
$$
\P\left(\sup_{1\le i \le m} |z'_i| \ge t\right) \le 2m \cdot  \phi(t)/t. 
$$
With $t = \sqrt{2 \log m}$, this gives $\P(\sup_{1\le i \le m} |z'_i|
\ge \sqrt{2\log m}) \le 1/\sqrt{\pi \log m}$. Better bounds are
possible but we will not pursue these refinements here. Observe now
that $s_i^2 = \|Q_{i,\cdot}\|^2_{\ell_2} = 1 -
\|A_{i,\cdot}\|^2_{\ell_2}$, and since $\lambda_i = \sqrt{2\log m}\,
\|Q_{i,\cdot}\|_{\ell_2}$, we have that 
\begin{equation}
  \label{eq:event}
  |QQ^* z_i| \le \lambda_i, \quad \forall i
\end{equation}
with probability at least $1-1/\sqrt{\pi \log m}$.

On the event \eqref{eq:event}, Theorem \ref{teo:Pinftydecode}
then shows that
\begin{equation}
  \label{eq:p1decodeused}
  \|\hat x - x\|_{\ell_2} \le C \, \sqrt{k} \cdot (m/r) \cdot \max_i 
|\lambda_i| + \|x - x^{\text{Ideal}}\|_{\ell_2}.
\end{equation}
We claim that
\begin{equation}
  \label{eq:normsQ}
  \frac{\max_i |\lambda_i|}{\sqrt{2\log m}} = 
\max_i \|Q_{i,\cdot}\|_{\ell_2} \le \sqrt{\frac{3r}{m}}
\end{equation}
with probability at least $1-2e^{-\gamma m}$ for some positive
constant $\gamma$.  Combining \eqref{eq:p1decodeused} and
\eqref{eq:normsQ} yields 
\[
\|\hat x - x\|_{\ell_2} \le 2C \cdot \sqrt{\frac{m \log m}{m-n}} \cdot
\sqrt{k} + \|x - x^{\text{Ideal}}\|_{\ell_2}.
\]
This would essentially conclude the proof of the corollary since the
size of $\|x - x^{\text{Ideal}}\|_{\ell_2}$ is about
$\sqrt{n}$. Exact bounds for $\|x - x^{\text{Ideal}}\|_{\ell_2}$
are found in the proof of Corollary \ref{cor:P2decode} and we do not
repeat the argument. 

It remains to check why \eqref{eq:normsQ} is true. For $r \ge m/3$ and
since $\|Q_{i,\cdot}\|_{\ell_2} \le 1$, the claim holds with
probability 1 because $3r/m \ge 1$! For $r \le m/3$, it follows from
$\|Q_{i,\cdot}\|^2_{\ell_2} + \|A_{i,\cdot}\|^2_{\ell_2} = 1$ that
\[
\P\left(\max_i \|Q_{i,\cdot}\|^2_{\ell_2} \ge \frac{2r}{m}\right) =
\P\left(\min_i \|A_{i,\cdot}\|^2_{\ell_2} \le
  \frac{n}{m}\left(1-\frac{r}{n}\right)\right) \le m
\, \P\left(\|A_{1,\cdot}\|^2_{\ell_2} \le
  \frac{n}{m}\left(1-\frac{r}{n}\right)\right). 
\]
The claim follows by applying Lemma \ref{teo:norms} since $r/n \le
1/2$.

\section{Discussion}
\label{sec:discussion}

We have introduced two decoding strategies for recovering a block $x
\in \R^n$ of $n$ pieces of information from a codeword $A x$ which has
been corrupted both by adversary and small errors.  Our methods are
concrete, efficient and guaranteed to perform well. Because we are
working with real valued inputs, we emphasize that this work has
nothing to do with the use of linear programming methods proposed by
Feldman and his colleagues to decode binary codes such as turbo-codes
or low-density parity check codes
\cite{Feldman1,Feldman2,Feldman3}. Instead, it has much to do with the
recent literature on compressive sampling or compressed sensing
\cite{CRT,OptimalRecovery,donoho04co,TroppGilbert,CDDCS,VershyninRudelson},
see also \cite{Vetterli1,Vetterli2} for related work.

On the practical end, we truly recommend using the two-step refinement
discussed in Section \ref{sec:simulations}---the reprojection
step---as this really tends to enhance the performance. We anticipate
that other tweaks of this kind might also work and provide additional
enhancement. On the theoretical end, we have not tried to obtain the
best possible constants and there is little doubt that a more careful
analysis will provide sharper constants. Also, we presented some
results for coding matrices with orthonormal columns for ease of
exposition but this is unessential. In fact, our results can be
extended to nonorthogonal matrices. For instance, one could just as
well obtain similar results for $m \times n$ coding matrices $A$ with
independent Gaussian entries.

There are also variations on how one might want to decode. We focused
on constraints of the form $\|P_{V^\perp} \tilde z\|$ where $\|\cdot
\|$ is either the $\ell_2$ norm or the $\ell_\infty$ norm, and
$P_{V^\perp}$ is the orthoprojector onto $V^\perp$, the orthogonal
subspace to the column space of $A$. But one could also imagine
choosing other types of constraints, e.g.~of the form $\|X^*\tilde
z\|_{\ell_2} \le \eps$ for $(P_2)$ or $\|XX^* \tilde z\|_{\ell_\infty}
\le \lambda$ for $(P_\infty)$ (or constraints about the individual
magnitudes of the coordinates $(XX^* \tilde z)_i$ in the more general
formulation), where the columns of $X$ span $V^\perp$. In fact, one
could choose the decoding matrix $X$ {\em first}, and then $A$ so that
the ranges of $A$ and $X$ are orthogonal. Choosing $X \in R^{m \times
  r}$ with i.i.d.~mean-zero Gaussian entries and applying the LP
decoder with a constraint on $\|XX^* \tilde z\|_{\ell_\infty}$ instead
of $\|\tilde z\|_{\ell_\infty}$ would simplify the argument since
restricted isometry constants for Gaussian matrices are already
readily available
\cite{OptimalRecovery,DecodingLP,Equivl0l1,rudelson06sp}!

Finally, we discussed the use of coding matrices which have fast
algorithms, thus enabling large scale problems. Exploring further
opportunities in this area seems a worthy pursuit.

\section{Appendix: Proof of Lemma \ref{teo:DS}}
\label{sec:appendix}

  The proof is adapted from that of Lemma 3.1 in \cite{Dantzig}. In
  the sequel, $T_0 \subset \{1, \ldots, m\}$ is a set of size $k$,
  $T_1$ is the $k'$ largest positions of $h$ outside of $T_0$, $T_{01}
  = T_0 \cup T_1$ and $V_{01} \subset \R^m$ is the subspace spanned by
  the columns of $\Phi$ with indices in $T_{01}$. Below, we omit the
  dependence on $\Phi$ in the constants $a_{k}(\Phi)$ and
  $\theta_{k,k'}(\Phi)$.

  Let $P_{V_{01}}$ be the orthogonal projection onto ${V_{01}}$. For each $w \in
  \R^m$, $\Phi_{T_{01}}^* w = \Phi_{T_{01}}^* P_{V_{01}} w$ and thus, 
\begin{equation}
  \label{eq:proj}
 \|\Phi^*_{T_{01}} w \|_{\ell_2} 
  \geq a_{k+k'} \,  \| P_{V_{01}} w \|_{\ell_2}, 
\end{equation}
since all the singular values of $\Phi_{T_{01}}$ are lower bounded by
$a_{|T_{01}|} = a_{k+k'}$.  With $w = \Phi h$, this gives
\begin{equation}
  \label{pivah}
  \| P_{V_{01}} \Phi h \|_{\ell_2} \leq a_{k+k'}^{-1} \|
  \Phi_{T_{01}}^* \Phi h \|_{\ell_2}.
\end{equation}

Next, divide $T^c_0$ into subsets of size $k'$ and enumerate $T^c_0$
as $n_1,n_2,\ldots,n_{m-|T_0|}$ in decreasing order of magnitude of
$h_{T^c_0}$.  Set $T_j = \{n_\ell, (j-1)k' + 1 \leq \ell \leq j k'\}$.
That is, $T_1$ is as before and contains the indices of the $k'$
largest coefficients of $h_{T^c_0}$, $T_2$ contains the indices of the
next $k'$ largest coefficients, and so on. We will develop a lower
bound on the $\ell_2$ norm of $P_{V_{01}} \Phi h$, which we
decompose as
\begin{equation}\label{pisplat}
  P_{V_{01}} \Phi h = P_{V_{01}} \Phi h_{T_{01}} + \sum_{j\ge2} 
P_{V_{01}} \Phi h_{T_j} = \Phi h_{T_{01}} + \sum_{j\ge2} P_{V_{01}} \Phi h_{T_j}. 
\end{equation}
By definition $P_{V_{01}} \Phi h_{T_j} \in {V_{01}}$ and thus 
\[
P_{V_{01}} \Phi h_{T_j} = \Phi_{T_{01}} c \quad \Rightarrow \quad a_{k+k'}
\|c\|_{\ell_2} \le \|P_{V_{01}} \Phi h_{T_j}\|_{\ell_2}, 
\]
which again follows from the lower bound on the singular values of
$\Phi_{T_{01}}$ (the coefficient sequence $c$ depends on $j$).
Observe now that
\begin{align*}
  \| P_{V_{01}} \Phi h_{T_j} \|_{\ell_2}^2 = \langle P_{V_{01}} \Phi h_{T_j}, \Phi
  h_{T_j} \rangle
  & =  \langle \Phi_{T_{01}} c , \Phi h_{T_j} \rangle\\
  & \le \theta_{k+k',k'} \, \|c\|_{\ell_2} \, \| h_{T_j} \|_{\ell_2}
  \le \frac{\theta_{k+k',k'}}{a_{k+k'}} \, \| P_{V_{01}} \Phi h_{T_j}
  \|_{\ell_2} \, \|h_{T_j}\|_{\ell_2},
\end{align*}
where the first inequality follows from the definition \eqref{ortho}
of the number of $\theta_{k+k',k'}$. In short,
\begin{equation}
  \label{eq:good}
  \| P_{V_{01}} \Phi h_{T_j} \|_{\ell_2}  \leq  \frac{\theta_{k+k',k'}}{a_{k+k'}}  
\, \| h_{T_j} \|_{\ell_2}.
\end{equation}
We then develop an upper bound on $\sum_{j \ge 2} \| h_{T_j} \|_{\ell_2}$
as in \cite{CRT2}.  By construction, the magnitude of each coefficient
in $T_{j+1}$ is less than the average of the magnitudes in $T_j$, 
\[
\|h_{T_{j+1}}\|_{\ell_\infty} ~\leq \|h_{T_j}\|_{\ell_1}/k' \quad
\Rightarrow \quad \|h_{T_{j+1}}\|^2_{\ell_2} \leq
\|h_{T_j}\|^2_{\ell_1}/k'.
\]
Therefore, 
\begin{equation}
  \label{eq:coding}
  \sum_{j\geq 2} \|h_{T_j}\|_{\ell_2} \le \sum_{j\geq 1}
  \|h_{T_j}\|_{\ell_1}/\sqrt{k'} = \|h\|_{\ell_1(T_0^c)}/\sqrt{k'}. 
\end{equation}
Hence, we deduce from \eqref{pisplat} that 
\begin{align*}
  \| P_{V_{01}} \Phi h\|_{\ell_2} \geq \|\Phi h_{T_{01}}\|_{\ell_2} -
  \sum_{j\ge2} \|P_{V_{01}} \Phi h_{T_j}\|_{\ell_2}
  & \ge a_{k+k'} \|h_{T_{01}}\|_{\ell_2} -   \frac{\theta_{k+k',k'}}{a_{k+k'}}  \sum_{j\ge2} \| h_{T_j} \|_{\ell_2} \\
  & \ge a_{k+k'} \|h_{T_{01}}\|_{\ell_2} -
  \frac{\theta_{k+k',k'}}{a_{k+k'} \, \sqrt{k'}} \,
  \|h\|_{\ell_1(T_0^c)}.
\end{align*}
Combining this with \eqref{pivah} proves the first part of the lemma. 

For the second part, observe that the $j$th largest value of
$|h_{T^c_0}|$ obeys $|h_{T^c_0}|_{(j)} \le \|h_{T^c_0}\|_{\ell_1}/j$, whence
\[
\|h_{T^c_{01}}\|^2_{\ell_2} ~\leq ~ \|h_{T^c_0}\|^2_{\ell_1}
\sum_{j \ge k'+1} j^{-2} ~\leq ~\|h_{T^c_0}\|^2_{\ell_1}/k'.
\]
The lemma is proven.  

\small
\bibliographystyle{plain}
\bibliography{GrossErrorsSmallErrors}

\end{document}